\documentclass[12pt]{article}

\usepackage{amsfonts,amsmath,amsxtra}

\usepackage{graphicx}

\newcommand{\beq}[1]{\begin{equation}\label{#1}}
\newcommand\eeq {\end{equation}}
\newcommand\bqa {\begin{eqnarray}}
\newcommand\eqa {\end{eqnarray}}
\newcommand\pr {\partial}

\newcommand{\eq}[1]{eq.(\ref{#1})\ }
\newcommand{\bear}{\begin{array}}
\newcommand{\enar}{\end{array}}

\newcommand{\C}{\mathbb{C}}

\begin{document}
\def\t{\theta}
\def\T{\Theta}
\def\w{\omega}
\def\ov{\overline}
\def\a{\alpha}
\def\b{\beta}
\def\g{\gamma}
\def\s{\sigma}
\def\l{\lambda}
\def\wt{\widetilde}


\hfill{RIKEN-TH-44}

\hfill{ITEP-TH-39/05}

\vspace{10mm}

\centerline{\bf \Large Towards the Theory of Non--Abelian Tensor
Fields II}

\vspace{10mm}

\centerline{{\bf Emil T.Akhmedov
}\footnote{email:{akhmedov@itep.ru}}}

\vspace{5mm}

\centerline{117218, Moscow, B.Cheremushkinskaya, 25, ITEP, Russia}

\centerline{and}

\centerline{Theoretical Physics Laboratory, RIKEN, Wako 2-1,
Saitama, 351-0198,Japan}


\begin{abstract}
We go on with the definition of the theory of the non--Abelian
two--tensor fields and find the gauge transformation rules and
curvature tensor for them. To define the theory we use the surface
{\it exponent} proposed in hep--th/0503234. We derive the
differential equation for the {\it exponent} and make an attempt
to give a matrix model formulation for it. We discuss application
of our constructions to the Yang--Baxter equation for integrable
models and to the String Field Theory.

\end{abstract}

\vspace{10mm}

\section{Introduction}

It is well known that the standard theory of fiber bundles is
designed to deal with the situations when one has ambiguously
defined functions --- sections of fiber bundles over a space $X$.
In the modern language this means that, knowing the value of a
section of a fiber bundle at a point $x$, we can find its value at
any other point $y$. But the result will depend on the path taken
from $x$ to $y$ and from the choice of the one--form connection on
the bundle. The path ordered exponent of the connection gives the
map between the fibers $V_x$ and $V_y$ over the points $x$ and
$y$.

This construction assumes that the one--form connection is
unambiguously defined up to the gauge transformations ---
rotations in the fibers. But what if the one--form connection was
ambiguous?( The simplest example of such a situation, which we
know, is the Wu--Yang monopole.) To describe the situation with
the ambiguous one--form connections one has to have a
``connection'' for the connections
--- the one to be ordered over the surface inside
$X$. The latter orderings give 2D holonomies which describe maps
between the sequences of fibers over different curves, along which
the original one--form was ordered. One can consult e.g.
\cite{Baez:2004in}, \cite{Baez:2005sn} on the mathematics behind
such a construction.

One badly needs a local field theory description of the 2D
holonomies in terms of two--tensor fields. In
\cite{Akhmedov:2005tn} and here we propose such a theory. To get
it one has to organize a {\bf triangulation--independent}
area--ordering of two--tensor gauge field carrying color indices.
The triangulation--independence is very crucial for the
consistency of the theory. It means the following fact. To
construct the area--ordering we {\bf should} approximate the
Riemann surfaces by their triangulated versions. For the
discretized case the area ordering is obtained as follows. We glue
the exponents of the two--tensor field over the whole simplicial
surface by putting them at different simplices (triangles), which
are sitting at different points inside the base space $X$. Now it
is easy to see that the two--tensor connection should carry {\bf
three color} indices along with {\bf two spacial} ones
\cite{Akhmedov:2005tn}: $B_{\mu\nu}^{ijk}$, $\mu,\nu = 1, \dots,
{\rm dim} X$ and $i,j,k = 1,\dots, {\rm dim} V$. This is due to
the three wedges of each 2D simplex. In \cite{Akhmedov:2005tn} we
have found a way to {\it exponentiate} cubic matrices, which we
refer to as the surface {\it exponent}. With such an {\it
exponent} we can take the continuum limit from the simplicial
surfaces to the smooth ones. It is important that with the use of
the surface {\it exponent} the result for the limit will not
depend on the way it is taken! This is what we refer to as
triangulation--independence.

The organization of the paper is as follows. In the next section
we briefly review the paper \cite{Akhmedov:2005tn} and concisely
formulate its statements. In the section 3 we find the
differential equation for the surface {\it exponent} and define a
new way of the {\it exponentiation} of the quadratic matrices. We
establish the relation between such an {\it exponent} of the
quadratic matrices and the surface {\it exponent} for the cubic
ones. At the end of the section 3 we discuss a possible matrix
integral formulation of the surface {\it exponent}. In the section
4 we derive the gauge transformation rules and curvature for the
two--tensor gauge fields. As well in that section we discuss the
meaning of the theory of the non--Abelian tensor fields. In the
section 5 we establish links between the latter theory and the
lattice integrable models and String Field Theory. We conclude
with summary in the section 6.

\section{{\it Exponentiation} of cubic matrices and the area--ordering}

\subsection{The surface {\it exponent}}

 In the paper \cite{Akhmedov:2005tn} we have defined the surface {\it exponent}
of a cubic matrix $\hat{B} = ||B_{ijk}||$, $i,j,k = 1,...,N$ as
follows\footnote{The reason for the subscripts on the LHS of this
equation will be explained below.}:

\bqa {\rm Tr}\, \left( E^{\hat{B}}_{g,I, \kappa}\right) \equiv
\lim_{M\to\infty} \prod_{{\rm graph}, g}^M \left(\hat{I} +
\frac{\hat{B}}{M}\right),\label{def}\eqa where the limit is taken
over a sequence of {\bf closed} (because on the LHS we take Tr
--- no any free indices), connected three--valent
graphs\footnote{Three--valent graphs are those in which three
wedges terminate in each their vertex. Note that in
\cite{Akhmedov:2005tn} we have considered the triangulation graphs
--- those which triangulate Riemann surfaces and are dual to the {\bf
fat} three-valent graphs (see fig. \ref{dual}).} with $M$
vertices. At a fixed $M$ the product on the RHS of \eq{def} is
taken over an $M$--vertex three--valent graph: At each vertex of
the graph we put matrix $I_{ijk} + B_{ijk}/M$ and we glue the
indices of these matrices with the use of a bilinear form
$\kappa^{ij}$ (as it is shown in the fig. \ref{pyramid}). It seem
that to take the limit $M\to\infty$ we have to choose a sequence
of graphs. But in \cite{Akhmedov:2005tn} we have proved that {\bf
the limit (\ref{def}) does NOT depend on the choice of the
sequence of graphs} under the following conditions:

\begin{figure}
\begin{center}
\includegraphics[scale=0.5]{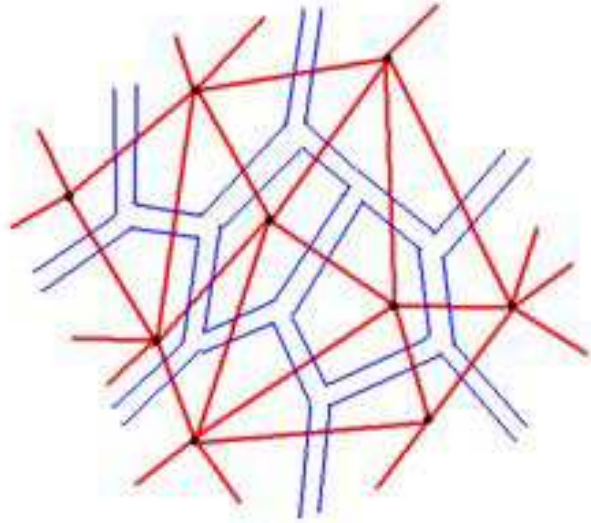}\caption{\footnotesize The duality relation
between graphs. The three--valent graph is shown here by the {\bf
fat} stripes. The relation between the dual graphs is as follows.
We place the vertices of the dual graph at the centers of the
faces of the original one and join the vertices via the wedges of
the dual graph passing through the wedges of the original one.
Thus, the dual graph to a {\bf fat} three--valent one is the
triangulation graph
--- the graph whose faces are triangles. In this note we are
actually using {\bf fat} three--valent graphs and are frequently
changing between three--valent and triangulation
graphs.}\label{dual}
\end{center}
\end{figure}

\begin{figure}
\begin{center}
\includegraphics[scale=0.5]{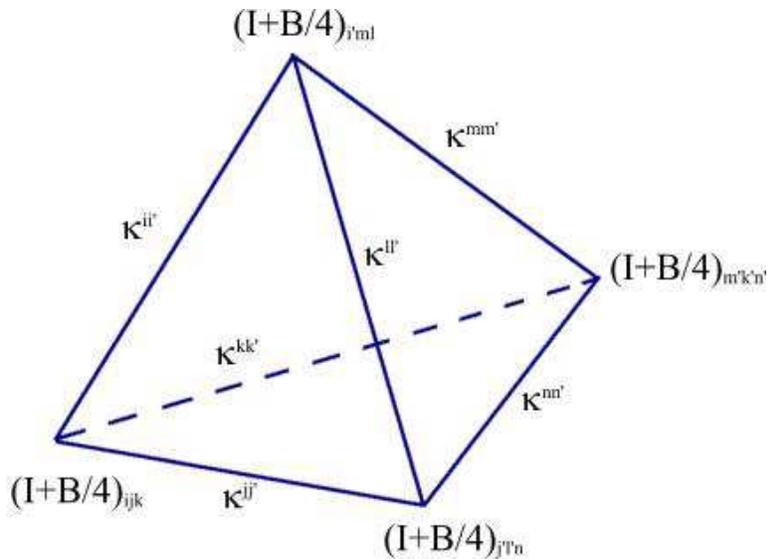}\caption{\footnotesize The example of the product of
the matrix $(\hat{I}+\hat{B}/4)$ over the pyramid, i.e. when
$M=4$.}\label{pyramid}
\end{center}
\end{figure}

\begin{itemize}

\item One should consider {\bf fat} three--valent graphs so that
it is obvious on which minimal genus Riemann surface this graph
can be mapped. The genus $g$ of the graphs in the sequence should
be fixed: I.e. at every $M$ the same topology graphs should be
taken.

\item The matrices $\hat{I}$ and $\hat{\kappa}$ should obey the
following conditions:

\bqa I_{ijk} = I_{kij} = I_{jki} - {\rm cyclic \,\, symmetry},\nonumber \\
\sum_{j,k=1}^N {I_{ij}}^k \, {I_{lk}}^j = \kappa_{il} - {\rm
normalization}, \nonumber
\\ \sum_{n=1}^N I_{inl} \, {I^n}_{jk} = \sum_{n=1}^N I_{ijn} \, {I^n}_{kl} - {\rm fusion}.
\label{cond}\eqa and we higher the indices of $I_{ijk}$ with the
use of $\kappa^{ij}$. One can find the graphical representation of
the last two equations in the fig. \ref{konditions}.

\begin{figure}
\begin{center}
\includegraphics[scale=0.5]{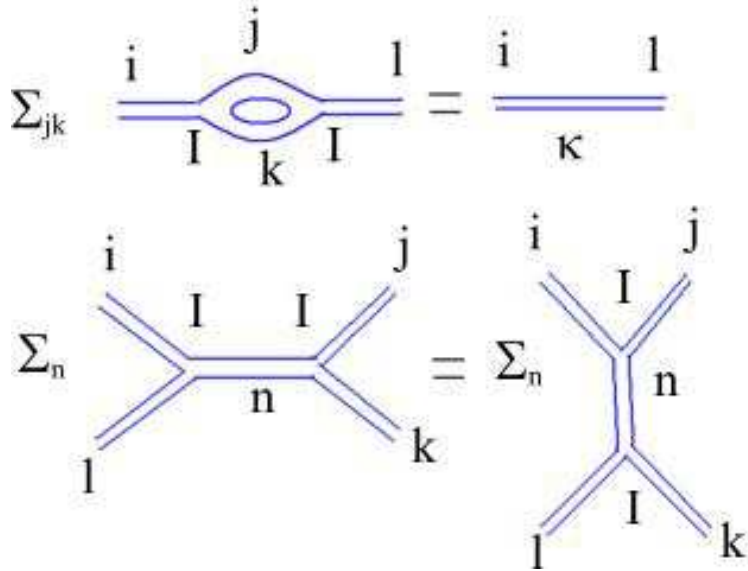}\caption{\footnotesize The two--dimensional
conditions which are related to the triangulation independence.
These conditions are solved via the use of a {\bf semi--simple
associative} algebra (see e.g. \cite{Verlinde:1988sn},
\cite{Fukuma:1992hy}, \cite{Akhmedov:2005tn}): ${I_{ij}}^k =
I_{ijl}\, \kappa^{lk}$ are structure constants of this algebra,
while $\kappa_{ij}$ is the non--degenerate Killing form. If the
algebra is commutative then the matrix $I_{ijk}$ can be mapped
(via an orthogonal rotation) to $\delta_{ijk}$ --- cubic matrix
whose only non--zero elements are units standing on the main
diagonal of the cube. In this case the surface {\it exponent} can
be mapped to the standard one dependent only on the diagonal
components of the matrix $\hat{B}$ \cite{Akhmedov:2005tn}. The
non--trivial case appears when the underlaying semi--simple
algebra is non--commutative. The first such case appears when
$N=4$ and corresponds to the algebra of $2\times 2$ matrices.
}\label{konditions}
\end{center}
\end{figure}

\item In the sequence we confine to the graphs of the following
kind: As the limit $M\to\infty$ is taken the number of wedges of
each face of the graph should be suppressed in comparison with
$M$. The geometric meaning of this condition is explained in
\cite{Akhmedov:2005tn}.

\end{itemize}
In \cite{Akhmedov:2005tn} we consider one extra condition.
However, this condition is not necessary for the limit (\ref{def})
to be independent of the choice of the sequence of the graphs. It
just simplifies the considerations of \cite{Akhmedov:2005tn}.

\begin{figure}
\begin{center}
\includegraphics[scale=0.5]{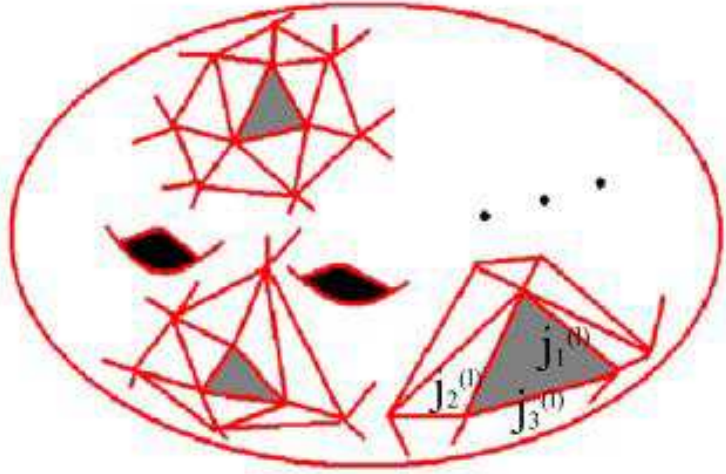}\caption{\footnotesize The dual to the three--valent
(triangulation) graph defining
$I^g_{j^{(1)}_3\,j^{(1)}_2\,j^{(1)}_1\left| \dots \right|
j^{(F)}_3\,j^{(F)}_2\,j^{(F)}_1}$.}\label{Ig}
\end{center}
\end{figure}

In general at given $N$ there are many solutions to the equations
for $\hat{I}$ and $\hat{\kappa}$ and the {\it exponent} depends on
the choice of $\hat{I}$, $\hat{\kappa}$ and $g$. This explains the
subscripts in \eq{def}. The result of the limit is:

\bqa {\rm Tr} \, \left(E^{\hat{B}}_{g,I,\kappa}\right) =
\sum_{F=0}^{\infty} \frac{1}{F!} \,
B^{j^{(1)}_1\,j^{(1)}_2\,j^{(1)}_3} \dots
B^{j^{(F)}_1\,j^{(F)}_2\,j^{(F)}_3} \,
I^g_{j^{(1)}_3\,j^{(1)}_2\,j^{(1)}_1\left| \dots \right|
j^{(F)}_3\,j^{(F)}_2\,j^{(F)}_1}, \label{defexp}\eqa where we use
the same letter $I$ to denote another matrix
$I^g_{j^{(1)}_3\,j^{(1)}_2\,j^{(1)}_1\left| \dots \right|
j^{(F)}_3\,j^{(F)}_2\,j^{(F)}_1}$ with $3F$ indices, which is
obtained via multiplication of $I_{ijk}$ over {\bf any} oriented
three--valent graph with $g$ handles, $F$ holes and 3 external
wedges at each hole (see fig. \ref{Ig}). At the same time $I^g \,
(F=0)$ is just a number obtained via the multiplication of
$I_{ijk}$ over any closed genus $g$ graph. For example, $I^0 =
\sum_{i,k} \kappa_i^k \, \kappa_k^i = \sum_i \delta_i^i = N$. The
reason for the division of the indices of
$I^g_{j^{(1)}_3\,j^{(1)}_2\,j^{(1)}_1\left| \dots \right|
j^{(F)}_3\,j^{(F)}_2\,j^{(F)}_1}$ into groups separated by
vertical lines is as follows: These matrices are only {\bf
cyclicly} symmetric under the exchange of their indices inside
each triple, but are {\bf completely} symmetric under any exchange
of the triples between themselves \cite{Akhmedov:2005tn}. All that
follows from the topology of the underlaying graphs (see fig.
\ref{Ig}) and the condition in fig. \ref{konditions}.

Once we have defined the trace of the surface {\it exponent}, we
can define the {\it exponent} with any number of indices or,
better to say, for any two--dimensional topology --- for a given
number of handles $g$, holes $L$ and distribution of external
indices over the holes. To begin with, let us define the {\it
exponent} for the disc topology and three external wedges at the
boundary. To do that in \eq{def} we have to use open graph with
one hole and three wedges at the boundary. The result for the
corresponding limit is:

\bqa \left(E^{\hat{B}}_{I,\kappa}\right)_{m_1 \, m_2\, m_3} =
\sum_{F=0}^{\infty} \frac{1}{F!} \,
B^{j^{(1)}_1\,j^{(1)}_2\,j^{(1)}_3} \dots
B^{j^{(F)}_1\,j^{(F)}_2\,j^{(F)}_3} \, I_{\left. m_1 \, m_2 \, m_3
\right| j^{(1)}_3\,j^{(1)}_2\,j^{(1)}_1\left| \dots \right|
j^{(F)}_3\,j^{(F)}_2\,j^{(F)}_1}.\label{builbl} \eqa  The {\it
exponent} in \eq{builbl} is the building block for the
construction of a generic surface {\it exponent} and for the
area--ordering below. In fact, we can glue a two--dimensional
surface of any topology with the use of triangles --- dual to the
three--valent vertices. Now we have to put
$\left(E^{\hat{B}}_{I,\kappa}\right)_{m_1 \, m_2\, m_3}$ instead
of $(I+B/M)_{ijk}$ or $I_{ijk}$ in the vertices. For example,
\eq{defexp} is obtained via multiplication of \eq{builbl} over any
genus $g$ closed triangulated Riemann surface: To obtain
Tr$E^{\hat{B}}$ we have to take $\hat{B}/J$ in \eq{builbl}, where
$J$ is the total number of triangles out of which the closed
surface in question is constructed. This is true due to the main
property of the surface {\it exponent} which is discussed in the
section 3.

\subsection{Triangulation--independent area--ordering}

\begin{figure}
\begin{center}
\includegraphics[scale=0.5]{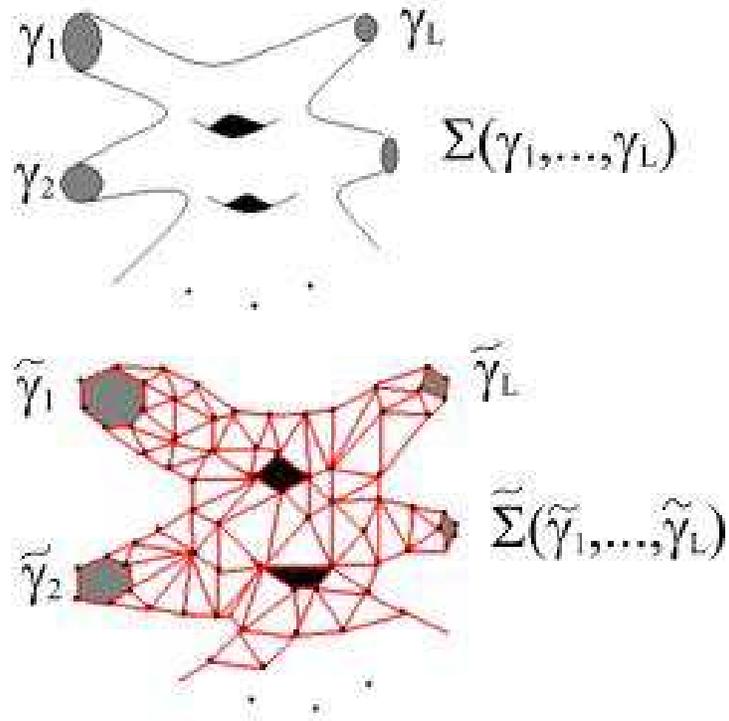}\caption{\footnotesize $\tilde{\Sigma}$
is the discretization of $\Sigma$; $\tilde{\gamma}$'s are the
discretizations of $\gamma$'s.}\label{fig3}
\end{center}
\end{figure}

With the use of the surface {\it exponent}, in
\cite{Akhmedov:2005tn} we have constructed the {\bf
triangulation--independent} ordering of non--Abelian two--tensor
fields over two--dimensional surfaces. Let us repeat that
construction here. It is natural to consider, within this context,
a triangulated approximation $\tilde{\Sigma}(\tilde{\gamma_1},
\dots, \tilde{\gamma_L})$ of an oriented Riemann surface
$\Sigma(\gamma_1, \dots, \gamma_L)$ in a target space $X$. This
surface has $L$ boundary closed loops $\gamma$'s which are
approximated by the closed broken lines $\tilde{\gamma}$'s (see
fig. \ref{fig3}). At the end we take the continuum limit and the
result does not depend on the way the limit is taken: It does not
depend on the choice of the sequence of triangulated surfaces
$\tilde{\Sigma}$ which are approaching the smooth surface $\Sigma$
in the limit in question! It is this fact which we refer to as the
triangulation independence.

To get the {\bf triangulation--independent} area--ordering, we
have to assign the matrix

$$U_{ijk}(x,\, \Delta\sigma^{\mu\nu}) = \left(E_{I,\kappa}^{\hat{B}_{\mu\nu}(x) \,
\Delta x^\mu\, \Delta x^\nu} \right)_{ijk}$$ to each simplex
(triangle) sitting at the point $x$ of the image of the
triangulated surface in $X$. Here $\Delta\sigma^{\mu\nu} = \Delta
x^\mu\, \Delta x^\nu$ is the oriented area element associated with
the triangle in question. The indices $i, j$ and $k$ are assigned
to the three wedges of the corresponding triangle. Hence,
$\hat{U}: V^3 \to \C $ for an $N$--dimensional vector space $V$
--- the ``fiber'' of our "fiber bundle", while $X$ is the base of the bundle.

\begin{figure}
\begin{center}
\includegraphics[scale=0.5]{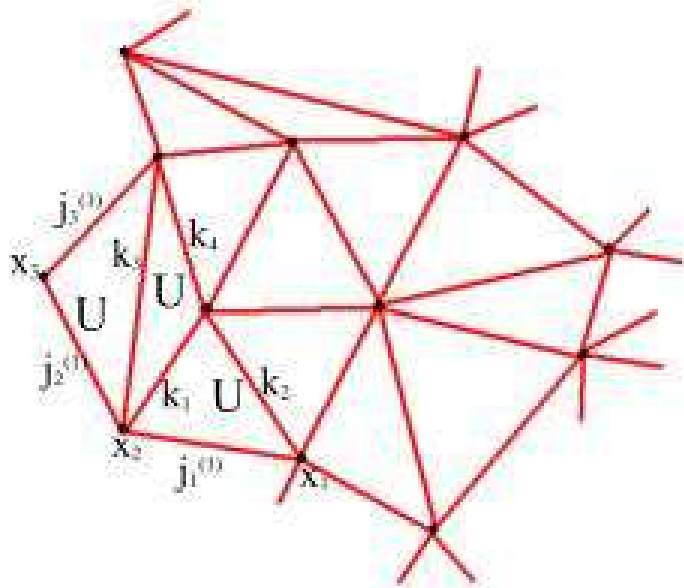}\caption{\footnotesize This is a fragment of
$\tilde{\Sigma}(\tilde{\gamma}_1,\dots, \tilde{\gamma}_L)$. On
each triangle of this figure there is $U$ matrix with three
indices and we sum over the indices assigned to the internal
wedges of the graph.}\label{fig4}
\end{center}
\end{figure}

 The area--ordering is
obtained by gluing, via the use of the bi--linear form
$\kappa^{ij}: \C \to V^2$, the matrices $U_{ijk}$ on each triangle
over the whole simplicial surface (see fig. \ref{fig4}):

\bqa U\left[\tilde{\Sigma}(\tilde{\gamma_1}, \dots,
\tilde{\gamma_L})\right]_{j^{(1)}_1\dots j^{(1)}_{n_1}\left|
j^{(2)}_1\dots j^{(2)}_{n_2}\right| \dots \left| j^{(L)}_1 \dots
j^{(L)}_{n_L}\right.} \equiv \nonumber \\ \equiv \sum_{k_1, k_2,
\dots, k_w} U_{j^{(1)}_1\, k_1\, k_2}
\left(x_1^{\phantom{\frac12}}, \Delta\sigma_1\right) \, {U^{k_1 \,
k_3}}_{k_4} \left(x_2^{\phantom{\frac12}}, \Delta\sigma_2\right)\,
{U^{k_3}}_{j^{(1)}_2\, j^{(1)}_3} \left(x_3^{\phantom{\frac12}},
\Delta\sigma_3\right)\dots \label{areaod}\eqa where $w$ is the
total number of internal wedges of the graph; $j^{(l)}$ are the
indices corresponding to the wedges of the $l$-th broken line
$\tilde{\gamma}_l$ and $n_l$ is the total number of the wedges of
this broken line. We higher (lower) the indices via the use of the
aforementioned bilinear form $\kappa^{ij}$ ($\kappa_{ij}
\kappa^{jk} = \delta_i^k$).

  In the continuum limit the discrete indices $1, ..., n_l$ are converted into the
continuum ones $s_l \in [0,\, 2\pi)$ and we obtain:

\bqa U\left[\tilde{\Sigma}(\tilde{\gamma_1}, \dots,
\tilde{\gamma_L})\right]_{j^{(1)}_1\dots j^{(1)}_{n_1}\left|
j^{(2)}_1\dots j^{(2)}_{n_2}\right| \dots \left| j^{(L)}_1 \dots
j^{(L)}_{n_L}\right.} \longrightarrow
U\left[\Sigma^{\phantom{\frac12}}(\gamma_1, \dots,
\gamma_L)\right]_{j^{(1)}(s_1)\left| j^{(2)}(s_2)\right| \dots
\left| j^{(L)}(s_L)\right.}, \nonumber \\ {\rm where} \quad
\hat{U}(\Sigma): V^{\infty}(1) \otimes \dots \otimes V^{\infty}(L)
\to \C. \label{cont}\eqa Here $j^{(l)}(s_l)$ is the ``color''
index assigned to the continuous number of points enumerated by
$s_l$
--- a parametrization of the $l$-th loop $\gamma_l$.
The expression (\ref{cont}) is not so unfamiliar for the string
theoreticians as it could seem from the first sight. In fact, if
we substitute $j^{(l)}(s_l)$ by $x^{(l)}(s_l)$, then
$U\left[\Sigma(\gamma_1, \dots, \gamma_L)\right]$ can be
considered as a kind of string amplitude whose end--loops
$\gamma$'s are mapped by $x(s)$'s. This is the subject of the
section 5.

 Using the surface {\it exponent} (\ref{builbl}), we can write the explicit
expression for the area--ordered {\it exponent} (``$A\,E$'')
(\ref{areaod})--(\ref{cont}). For example, for the case of the
disc $D$ we obtain:

\bqa U_{j(s)}(D, \hat{B}) = \left(A \,
E_{g,I,\kappa}^{{\int\int}_{D} \hat{B}_{\mu\nu} \, dx_\mu \,
dx_\nu}\right)_{j(s)} \equiv \sum_{F=0}^{\infty} \frac{1}{F!} \,
I^g_{\left. j(s)\right| j^{(1)}_3\,j^{(1)}_2\,j^{(1)}_1\left|
\dots \right| j^{(F)}_3\,j^{(F)}_2\,j^{(F)}_1} \, \times \nonumber
\\ \times {\int\int}_{D} B_{\mu_1\nu_1}^{j^{(1)}_1\,j^{(1)}_2\,j^{(1)}_3}
dx_{\mu_1} \, dx_{\nu_1} \dots {\int\int}_{D}
B_{\mu_F\nu_F}^{j^{(F)}_1\,j^{(F)}_2\,j^{(F)}_3} dx_{\mu_F} \,
dx_{\nu_F} = \left(E_{g,I,\kappa}^{{\int\int}_{D} \hat{B}_{\mu\nu}
\, dx_\mu \, dx_\nu}\right)_{j(s)},\label{areaod1} \eqa where $s$
is the parametrization of the boundary and $j(s)$ is the index
function at the boundary $\pr D$. The last equality holds because
of the symmetry properties of the matrices $I_{\left. j(s) \right|
j^{(1)}_3\,j^{(1)}_2\,j^{(1)}_1\left| \dots \right|
j^{(F)}_3\,j^{(F)}_2\,j^{(F)}_1}$. These properties are discussed
below the \eq{defexp}. As the result, such an area--ordering is
rather trivial because there is no need to order anything. In
fact, we can easily interchange the order of integrals over $B$ in
\eq{areaod1}. This should be the generic property of all ``higher
dimensional'' {\it exponents} (for multi-index matrices) due to
the triviality of the corresponding homotopy groups.

Let us clarify this important subject. In the one--dimensional
ordering for one--form gauge fields the order is important for the
following reason: If one throws away a point (insertion of a local
observable) from an open curve (along which the ordering is done)
it becomes disconnected. Then it is important what is on the left
and what is on the right from the point in question. On the other
hand, in the case of two--dimensional orderings, throwing away a
point is not sufficient to make the surface disconnected. Hence,
the order is unimportant. As the result there are no commutators
of {\bf local} fields which describe two--dimensional
area--orderings. We come back to this point again at the end of
the section 4.

However, the commutative nature of the area--ordering does not
mean that we have obtained something trivial! In fact, if we
consider a surface $\Sigma$ with one designated point $x$ and two
external wedges at this point the corresponding $U(\Sigma_x)_{ij}$
(or ${U(\Sigma_x)_i}^j = U(\Sigma_x)_{ik} \, \kappa^{kj}$) gives a
nontrivial (non--diagonal matrix) map. All that goes without
saying that more complicated matrices
${U(\Sigma_{x,y,\dots})_{ijm\dots}}^{kl\dots}$ give completely
non--trivial non--linear maps. Somehow the latter objects
intrinsically include interactions because the $B$--field and
especially the background $\hat{I}$ give non--linear maps
($\hat{B}, \hat{I}: V^3 \to \C$). As the result $U$'s give
non--trivial non--linear maps between the end--loops of the
corresponding surfaces.

In any case, to our mind this is a deep fact which is yet to be
understood to establish the relation of the subject in question to
the String Field Theory and to the standard theory of fiber
bundles. Note that in \cite{Akhmedov:2005zi} we describe how to
construct modified surface ``{\it exponent}'', which obeys less
trivial relations, and we think that the surface {\it exponent}
can appear in the applications in the latter form.

\section{Relevant properties of the {\it exponent}}

\subsection{Differential equation for the {\it exponent}}

\begin{figure}
\begin{center}
\includegraphics[scale=0.5]{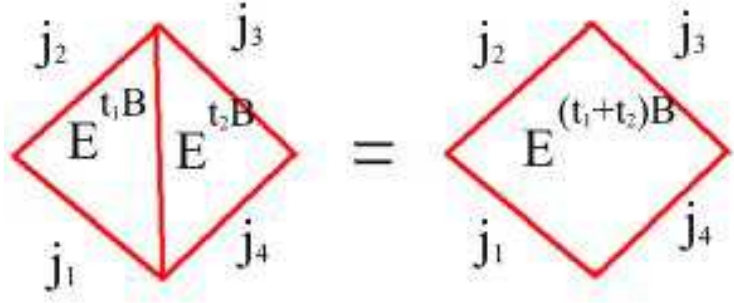}\caption{\footnotesize The basic condition
for the surface {\it exponent} shown for the example of the
triangulation graphs.}\label{diffeq}
\end{center}
\end{figure}

Let us derive the differential equation for the surface {\it
exponent}. To do that recall that the exponent of a quadratic
matrix ${A_i}^j$ has the following main property (which is at the
basis of the probability theory and all evolution phenomena):

\bqa {\left(e^{t_1 \, \hat{A}}\right)_i}^j \, {\left(e^{t_2 \,
\hat{A}}\right)_j}^k = {\left(e^{\left(t_1 + t_2\right)\,
\hat{A}}\right)_i}^k \label{diffeq1}\eqa for any two numbers $t_1$
and $t_2$. This equation defines the exponent unambiguously. In
fact, from \eq{diffeq1} one can derive the differential equation
for the exponent: Choose $t_1 = t$ and $t_2 = dt \ll t$. Then,
expanding over $dt$ both sides of \eq{diffeq1}, we obtain:

\bqa \frac{d}{dt} \left(e^{t\hat{A}}\right)_i^j = A_i^k
\left(e^{t\hat{A}}\right)_k^j.\eqa
 The surface {\it exponent} can have {\bf any} number of external indices --- not only zero
or two like the exponent of a quadratic matrix. As the result it
obeys many different identities following from the conditions of
the triangulation independence. But all these conditions originate
from the basic one\footnote{Note that according to the discussion
at the end of the section 2 we actually have the relation as:
${\left(E_{I,\kappa}^{\hat{B}_1}\right)_{j_1\, j_2}}^{k_1}
\,\left(E_{I,\kappa}^{\hat{B}_2}\right)_{j_3\, j_4 k_1} =
\left(E_{I,\kappa}^{\hat{B}_1 + \hat{B}_2}\right)_{j_1\, j_2\,
j_3\, j_4}$ for {\bf any} cubic matrices $\hat{B}_1$ and
$\hat{B}_2$.}:

\bqa {\left(E_{I,\kappa}^{t_1 \, \hat{B}}\right)_{j_1\,
j_2}}^{k_1} \,\left(E_{I,\kappa}^{t_2 \, \hat{B}}\right)_{j_3\,
j_4 k_1} = \left(E_{I,\kappa}^{\left(t_1 + t_2\right)\,
\hat{B}}\right)_{j_1\, j_2\, j_3\, j_4},\label{baric}\eqa which is
shown graphically in the fig. \ref{diffeq}. This equation,
however, does not define unambiguously the surface {\it exponent},
because $\hat{I}$ and $\hat{\kappa}$ do not explicitly present in
this equation.

However, if $t_1 = t$ and $t_2 = dt \ll t$, we can obtain the
following differential equation:

\bqa \frac{d}{dt}
{\left(E_{I,\kappa}^{t\,\hat{B}}\right)_{j_1\,j_2}}^{k_1}\,
I_{j_3\,j_4\,k_1} =
{\left(E_{I,\kappa}^{t\,\hat{B}}\right)_{j_1\,j_2}}^{k_1} \,
B_{j_3\,j_4\,k_1},\eqa which fixes the {\it exponent}
unambiguously for a particular choice of $\hat{I}$ and
$\hat{\kappa}$ matrices, because now they are explicitly present
in the equation.

Similarly one can write the variational equation for the
area--ordered exponent (\ref{areaod1}):

\bqa \frac{\delta^2}{\delta \sigma_{\mu\nu}^{2}(s)}
\left(E_{g,I,\kappa}^{{\int\int}_{D} \hat{B}_{\mu\nu} \, dx_\mu \,
dx_\nu}\right)_{\dots \, j_s} \, I^{j_skl} =
\left(E_{g,I,\kappa}^{{\int\int}_{D} \hat{B}_{\mu\nu} \, dx_\mu \,
dx_\nu}\right)_{\dots \, j_s}\,
B^{j_skl}_{\mu\nu}\left[x(s)\right], \eqa where $\delta^2/\delta
\sigma_{\mu\nu}^{2}(s)$ is the standard variation with respect to
the addition of a small area $(\Delta \sigma^{\mu\nu})$ to the
disc $D$ and $j_s$ means single index $j$ at the point $s$.

\subsection{New way to {\it exponentiate} quadratic matrices and its relation to
the surface {\it exponent}}

In this subsection we will define non--standard way of {\it
exponentiation} of {\bf quadratic} matrices. It is inspired by the
surface {\it exponent}. In fact, we can put in the vertices of the
graphs in \eq{def} the matrix $I_{ijk}$ rather than $I_{ijk} +
B_{ijk}/M$, but glue their indices with the use of $\kappa^{ij} +
B^{ij}/W$ rather than just with the use of $\kappa^{ij}$, i.e.:

\bqa{\rm Tr}\, \left( \tilde{E}^{\hat{B}}_{g,I,\kappa}\right)
\equiv \lim_{W\to\infty} \prod_{{\rm graph}(g)}^W
\left(\hat{\kappa} + \frac{\hat{B}}{W}\right),\label{def1}\eqa
where under ``graph'' we assume the same kind of three--valent
genus $g$ closed graph as in \eq{def}, which has $W$ wedges. In
the light of the above discussion the limit (\ref{def1}) does not
depend on the choice of the sequence of graphs. The result for the
limit is:

\bqa {\rm Tr} \, \left(\tilde{E}^{\hat{B}}_{g,I,\kappa}\right) =
\sum_{F=0}^{\infty} \frac{1}{F!} \, B^{j^{(1)}_1\,j^{(1)}_2} \dots
B^{j^{(F)}_1\,j^{(F)}_2} \, I^g_{j^{(1)}_1\,j^{(1)}_2\left| \dots
\right| j^{(F)}_1\,j^{(F)}_2},\eqa where
$I^g_{j^{(1)}_1\,j^{(1)}_2\left| \dots \right|
j^{(F)}_1\,j^{(F)}_2}$ is the matrix with $2F$ indices obtained
via multiplication of $I_{ijk}$ over {\bf any} oriented graph with
$g$ handles, $F$ holes and 2 external wedges at each hole.
Similarly we can define {\it exponents} with any distribution of
indices.

Now we are going to show that this {\it exponent} is equivalent to
the original one. In fact, it is easy to see that:

\bqa \lim_{W\to\infty} \prod_{{\rm graph}(g)}^W \left(\hat{\kappa}
+ \frac{\hat{B}}{W}\right) = \lim_{W\to\infty} \prod_{{\rm
graph}(g)}^W \left(\hat{\kappa} + \frac{\hat{B}}{2\,W}\right)^2 =
\lim_{M\to\infty} \prod_{{\rm graph}(g)}^M \left(1 +
\frac{\hat{B}}{3 \, M}\right)^3 \hat{I} \label{14},\eqa where $M$
is the number of vertices of the same graph. (Note that $2 \, W =
3\, M$ for the three--valent graphs.) In the last expression of
\eq{14} we put in the vertices the following matrix:

\bqa\left(1 + \frac{\hat{B}}{3 \, M}\right)^3 \hat{I} \equiv
\left|\left| \left(1 + \frac{B}{3 \, M}\right)^{i'}_i\left(1 +
\frac{B}{3 \, M}\right)^{j'}_j\left(1 + \frac{B}{3 \,
M}\right)^{k'}_k I_{i'j'k'}\right|\right| = \nonumber \\ =
\left|\left|I_{ijk} + \frac{B^{i'}_i}{3 \, M} \, I_{i'jk} +
\frac{B^{j'}_j}{3 \, M} \, I_{ij'k} + \frac{B^{k'}_k}{3 \, M} \,
I_{ijk'} + {\cal
O}\left(\frac{1}{M^2}\right)\right|\right|.\label{15}\eqa  We
higher the indices in this expression with the use of
$\kappa^{ij}$ and use in \eq{14} and \eq{15} the fact that
$\kappa_{ij}\, \kappa^{jk} = \delta_i^k$. In the limit $M\to
\infty$ the terms ${\cal O}(1/M^2)$ in the \eq{15} do not survive
and we obtain the relation between the cubic and quadratic
matrices:

\bqa \tilde{B}_{ijk} = \frac{B^{i'}_i}{3} \, I_{i'jk} +
\frac{B^{j'}_j}{3} \, I_{ij'k} + \frac{B^{k'}_k}{3} \,
I_{ijk'}\eqa for which the following equality holds:

\bqa\tilde{E}^{\hat{B}}_{g,I,\kappa} =
E^{\hat{\tilde{B}}}_{g,I,\kappa}.\eqa Thus, basically we have the
unique ``two--dimensional'' {\it exponent}. It is worth pointing
out now that all the relevant properties of the surface {\it
exponent} are due to the fact that it is the background $I_{ijk}$
who carries three color indices rather than the matrix $B_{ijk}$
under the exponent. In any case, we are going to use the new way
of {\it exponentiation} of the quadratic matrices in the section
4.

\subsection{Towards Matrix model representation for the {\it exponent}}

In this subsection we discuss our attempts to represent the
surface {\it exponent} via the use of a matrix integral. In fact,
the mentioned above matrices
$I^g_{j^{(1)}_3\,j^{(1)}_2\,j^{(1)}_1\left| \dots \right|
j^{(F)}_3\,j^{(F)}_2\,j^{(F)}_1}$ are 2D topological invariants in
the sense discussed say in \cite{Fukuma:1992hy}. This is true due
to the conditions (\ref{cond}) imposed on $\hat{I}$ and
$\hat{\kappa}$: The matrix
$I^g_{j^{(1)}_3\,j^{(1)}_2\,j^{(1)}_1\left| \dots \right|
j^{(F)}_3\,j^{(F)}_2\,j^{(F)}_1}$ depends only on the topology of
the discretized Riemann surface\footnote{By the ``topology'' in
this case we mean a given number of handles $g$, number of holes
$L$ and the distribution of external wedges over the holes.}, but
does not depend on its concrete triangulation. We would like to
write an explicit representation for this matrix through the
correlation function in a 2D topological theory.

To do that let us consider the matrix integral, which is a
generalization of the integral considered in
\cite{Kontsevich:1992ti}:

\bqa Z(K,\,N,\,\kappa,\, I) = \int \prod_{i=1}^N d\hat{\Phi}^i \,
\exp\left\{- K\, {\rm Tr} \left[\hat{\Phi}^i\,
\hat{\Lambda}\,\hat{\Phi}^j \, \kappa_{ij} + \frac{\rm i}{3}\,
I_{ijk} \, \hat{\Phi}^i \, \hat{\Phi}^j \,
\hat{\Phi}^k\right]\right\},\label{genKon}\eqa where from now on
we put $\hat{\Lambda} = 1$. The integral is taken over the unitary
matrices $\hat{\Phi}_i = ||\Phi_i^{ab}||$, where $i=1,...,N$ and
$a = 1,...,K$. Taylor expanding this integral over the powers of
$I_{ijk}$ and using the decoupling of correlations for the
Gaussian integral, usually referred to as Wick's theorem, we
obtain the Feynman diagram expressions for the contributions to
the integral. Then, it is easy to see that:

\bqa\left.\frac{\pr^{\chi(g)}}{\pr K^{\chi(g)}}
Z(K,\,N,\,\kappa,\, I)\right|_{K = 0} \propto I^g, \quad g \geq
1,\eqa because by taking this derivative we extract the genus $g$,
three--valent, closed graph (Feynman diagram), where in the
vertices the matrix $I_{ijk}$ is standing, while in the wedges one
puts the ``propagator'' $\kappa^{ij}$ ($\kappa^{ij}\,\kappa_{jk} =
\delta^i_k$). Note that the integral (\ref{genKon}) is convergent
and explicitly calculable:

\bqa \log Z(K,\,N,\,\kappa,\, I) = \sum_{g=0}^\infty C_g \,
K^{\chi(g)} \, I^g,\eqa where $C_g$ are some combinatorial
coefficients counting the numbers (with alternating signs) of
three--valent connected graphs at the given $g$.

It seems that by considering the $g$ handle contribution to the
correlation function:

\bqa\int \prod_{i=1}^N d\hat{\Phi}^i \,:\hat{\Phi}_{j_3^{(1)}} \,
\hat{\Phi}_{j_2^{(1)}} \, \hat{\Phi}_{j_1^{(1)}}: \dots
:\hat{\Phi}_{j_3^{(F)}} \, \hat{\Phi}_{j_2^{(F)}} \,
\hat{\Phi}_{j_1^{(F)}}: \, \exp\left\{- K\, {\rm Tr}
\left[\hat{\Phi}^i\, \hat{\Phi}^j \, \kappa_{ij} + \frac{\rm
i}{3}\, I_{ijk} \, \hat{\Phi}^i \, \hat{\Phi}^j \,
\hat{\Phi}^k\right]\right\}\label{corr}\eqa will give us the
desired matrix $I^g_{j^{(1)}_3\,j^{(1)}_2\,j^{(1)}_1\left| \dots
\right| j^{(F)}_3\,j^{(F)}_2\,j^{(F)}_1}$ if we do not Wick
contract the $\hat{\Phi}$'s inside the ``normal ordering'' ---
$:\Phi\,\Phi\,\Phi:$. In fact, this correlation function gives a
2D topological invariant in the sense of \cite{Fukuma:1992hy}.
However this topological invariant does not coincide  with
$I^g_{j^{(1)}_3\,j^{(1)}_2\,j^{(1)}_1\left| \dots \right|
j^{(F)}_3\,j^{(F)}_2\,j^{(F)}_1}$ because we can not separate the
contact terms from the desired ones: In the result of the
calculation of the genus $g$ contribution to the \eq{corr} (if
$F=2$), along with the matrix
$I^g_{\left.j^{(1)}_3\,j^{(1)}_2\,j^{(1)}_1\right|
j^{(2)}_3\,j^{(2)}_2\,j^{(2)}_1}$ the following kind of
contribution appears:
$I^g_{j^{(1)}_3\,j^{(1)}_2\,j^{(2)}_2\,j^{(2)}_1} \cdot
\kappa_{j^{(1)}_1\, j^{(2)}_3}$
--- one of the contact terms.

 At this stage we do not know how to separate this two kinds of
 contributions to the correlation functions (\ref{corr}). However,
 we would like to find a way to do that, because it will establish a closer
 relation of our considerations to the subject of the String Field
 Theory.

\section{Theory of non--Abelian two--tensor fields}

In this section we discuss gauge transformations and curvature for
the two--tensor gauge connection. To explain the idea of our
argument and to set the notations let us present here the
derivation of the gauge transformation and curvature of an
ordinary gauge field. Consider a holonomy matrix

$$\hat{U}(x, \, \Delta x) = e^{{\rm i}\,\hat{A}_\mu(x) \, \Delta x^\mu}$$
for a small $\Delta x$. This matrix transforms, under the
rotations in the fibers, as:

$$\hat{\widetilde{U}}(x, \, \Delta x) = \hat{g}^{-1}(x) \,
\hat{U}(x, \, \Delta x) \, \hat{g}(x + \Delta x).$$ Then,
expanding in powers of $\Delta x$ both sides of this expression,
we obtain:

$$\hat{\widetilde{A}}_\mu(x) = \hat{g}^{-1}(x) \,
\left[\pr_\mu + {\rm i}\,\hat{A}_\mu(x)\right]\, \hat{g}(x).$$
Which is the gauge transformation for the gauge field.

As well, by considering the holonomy matrix for a small square
loop and expanding in powers of its size, we obtain:

\bqa \hat{U}(x,\Delta x^\mu) \,\hat{U}(x + \Delta x^\mu, \Delta
x^\nu)\, \hat{U}^+(x + \Delta x^\nu, \Delta x^\mu)\, \hat{U}^+(x,
\Delta x^\nu) =  1 + {\rm i} \hat{F}_{\mu\nu}\, \Delta x^\mu \,
\Delta x ^\nu + \dots, \eqa where $\hat{F}_{\mu\nu} = \pr_{[\mu}
\, \hat{A}_{\nu]} - {\rm i} \,\left[\hat{A}_{\mu}, \,
\hat{A}_{\nu}\right]$.

Obviously for the two--form gauge connection everything works in a
similar way if we substitute the line holonomy for quadratic
matrices with the surface holonomy for cubic ones. For example, it
seems to be natural to postulate the following transformation for
the area--ordered {\it exponent} over the disc $D$:

\bqa \left(E_{I,\kappa}^{{\int\int}_{D}
\hat{\widetilde{B}}_{\mu\nu} \, dx_\mu \, dx_\nu}\right)_{j(s)} =
\left(E_{I,\kappa}^{{\int\int}_{D} \hat{B}_{\mu\nu} \, dx_\mu \,
dx_\nu}\right)_{k(s)} \,{\left( e^{{\rm i}\,\oint_{\pr D} ds \,
\dot{x}^\mu \, \hat{A}_\mu}\right)^{k(s)}}_{j(s)}, \eqa where $s$
is the parametrization of the boundary and $j(s)$ and $k(s)$ are
the index functions at the boundary $\pr D$. Here

\bqa{\left( e^{{\rm i}\,\oint_{\pr D} ds \, \dot{x}^\mu \,
\hat{A}_\mu}\right)^{k(s)}}_{j(s)} = \lim_{|\Delta x|\to 0,
L\to\infty}\prod_{a=1}^L {\left(e^{{\rm i}\, \hat{A}_{\mu}(x_a)
\Delta x^\mu}\right)^{k_a}}_{j_a}\eqa is a kind of a hedgehog line
--- ``Wilson line'' with all free (without contraction) indices of all
$\hat{A}$'s at each $s\in [0,2\pi)$. The exponent here is just the
standard one for quadratic matrices.

  However the $B$--field can not transform in this way. In fact, consider
such a transformation for a small square with four external
indices (see fig. \ref{square}):

\begin{figure}
\begin{center}
\includegraphics[scale=0.3]{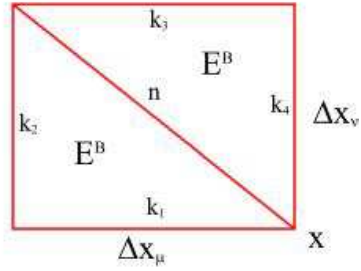}\caption{\footnotesize Disc topology with four external
wedges.}\label{square}
\end{center}
\end{figure}

\bqa\left(E_{I,\kappa}^{\hat{\widetilde{B}}_{\mu\nu}[x] \, \Delta
x^\mu \, \Delta x^\nu}\right)_{k_1\,k_2\,n} \,
{\left(E_{I,\kappa}^{ \hat{\widetilde{B}}_{\mu\nu}\left[x\right]
\, \Delta x^\mu \, \Delta x^\nu}\right)^n}_{k_3\,k_4} =
\left(E_{I,\kappa}^{ \hat{B}_{\mu\nu}[x] \, \Delta x^\mu
\, \Delta x^\nu}\right)_{j_1\,j_2 \,n} \times \nonumber \\
\times {\left(E_{I,\kappa}^{ \hat{B}_{\mu\nu}\left[x\right] \,
\Delta x^\mu \, \Delta x^\nu}\right)^n}_{j_3\,j_4} \,
{\left(e^{{\rm i}\,\hat{A}_\mu \Delta x^\mu}\right)^{j_1}}_{k_1}
\, {\left(e^{{\rm i}\,\hat{A}_\nu \Delta
x^\nu}\right)^{j_2}}_{k_2} \, {\left(\, e^{- {\rm i} \,
\hat{A}_\mu \Delta x^\mu}\right)^{j_3}}_{k_3} \, {\left(e^{- {\rm
i} \, \hat{A}_\nu \Delta x^\nu}\right)^{j_4}}_{k_4}.
\label{trans}\eqa Expanding this expression in powers of $\Delta
x$, we obtain:

\bqa  2\, I_{k_1\,k_2\,k_3\,k_4| m_3\,m_2\,m_1} \,
\widetilde{B}^{m_1\,m_2\,m_3}_{\mu\nu} \, \Delta x^\mu \, \Delta
x^\nu = 2\, I_{k_1\,k_2\,k_3\,k_4| m_3\,m_2\,m_1} \,
B^{m_1\,m_2\,m_3}_{\mu\nu} \, \Delta x^\mu \, \Delta x^\nu  + \nonumber \\
+ {\rm i} \, I_{j_1\,j_2\,j_3\,j_4}\,\left[\left(A_{\mu}\, \Delta
x^\mu\right)^{j_1}_{k_1}\, \delta^{j_2}_{k_2}\,
\delta^{j_3}_{k_3}\, \delta^{j_4}_{k_4} - \delta^{j_1}_{k_1}\,
\delta^{j_2}_{k_2}\, \left(A_{\mu}\, \Delta
x^\mu\right)^{j_3}_{k_3}\, \delta^{j_4}_{k_4}\right] + \nonumber
\\ + {\rm i} \, I_{j_1\,j_2\,j_3\,j_4}\,\left[\delta^{j_1}_{k_1}\,
\left(A_{\nu}\, \Delta x^\nu\right)^{j_2}_{k_2}\,
\delta^{j_3}_{k_3}\, \delta^{j_4}_{k_4} - \delta^{j_1}_{k_1}\,
\delta^{j_2}_{k_2}\, \delta^{j_3}_{k_3}\, \left(A_{\nu}\, \Delta
x^\nu\right)^{j_4}_{k_4}\right] + \dots \label{line}\eqa We do not
spell here the higher terms in $A$ because they are not relevant
to observe that there is no way to transform the $B$--field to
compensate the linear terms in $\Delta x$.

Then we propose that the area--ordered {\it exponent} transforms
with the use of the surface {\it exponent}:

\bqa\left(E_{I,\kappa}^{{\int\int}_{D}
\hat{\widetilde{B}}_{\mu\nu} \, dx_\mu \, dx_\nu}\right)_{j(s)} =
\left(E_{I,\kappa}^{{\int\int}_{D} \hat{B}_{\mu\nu} \, dx_\mu \,
dx_\nu}\right)_{k(s)} \,{\left( \tilde{E}_{I,\kappa}^{{\rm
i}\,\oint_{\pr D} ds \, \dot{x}^\mu \,
\hat{A}_\mu}\right)^{k(s)}}_{j(s)}, \eqa where we consider it more
natural to integrate the {\bf quadratic} matrix $\hat{A}_\mu =
||A^{ab}_\mu||$ (rather than a cubic one) over the boundary of the
disc $\pr D$. Hence in this expression we are using the {\it
exponent} defined in the subsection 3.2.

Now if we take the disc as in the fig. \ref{square} and expand
both sides of the corresponding expression (analogous to
\eq{trans}) in powers of $\Delta x$, we observe that all the
linear terms (as in \eq{line}) do cancel out and no other
dangerous terms appear. As the result the $B$--field transforms
as:

\bqa \tilde{B}^{ijk}_{\mu\nu} = B^{ijk}_{\mu\nu} + {\rm i} \,
\frac{{\pr_{[\mu} A_{\nu]}}^i_l}{3} \, I^{ljk} + {\rm i} \,
\frac{{\pr_{[\mu} A_{\nu]}}^j_l}{3} \, I^{ilk} + {\rm i} \,
\frac{{\pr_{[\mu} A_{\nu]}}^k_l}{3} \, I^{ijl}, \eqa i.e. as a
collection of the Abelian stringy $B$--fields for each group of
indices $i,j,k$. This is not a surprise for us once we understood
the main properties of the surface {\it exponent}, which are
discussed at the end of the section 2.

\begin{figure}
\begin{center}
\includegraphics[scale=0.3]{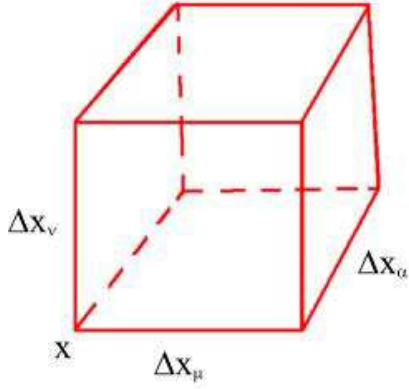}\caption{\footnotesize Holonomy matrix for cube gives the
curvature of the $B$--field.}\label{cube}
\end{center}
\end{figure}

We can calculate as well the curvature of the ``non--Abelian''
$B$--field. To do that we have to consider a small closed surface
having the geometry of the cube (see fig. \ref{cube}). Then the
corresponding holonomy matrix has the following expansion:

\bqa {\rm Tr} \, \left[\hat{U}\left(x, \Delta x^\mu \Delta
x^\nu\right) \hat{U}\left(x + \Delta x^\nu, \Delta x^\mu \Delta
x^\alpha\right) \, \hat{U}\left(x, \Delta x^\nu \Delta
x^\alpha\right) \right. \times \nonumber \\ \times \left.
\hat{U}^{-1}\left(x, \Delta x^\nu \Delta x^\mu\right)
\hat{U}^{-1}\left(x + \Delta x^\mu, \Delta x^\alpha \Delta
x^\nu\right) \, \hat{U}^{-1}\left(x + \Delta x^\alpha, \Delta
x^\nu \Delta x^\alpha\right)\right] = \nonumber
\\ = N + I_{ijk} \, \pr_{[\mu}B^{kji}_{\nu\alpha]} \Delta x^\mu
\Delta x^\nu \Delta x^\alpha + \dots, \label{YB}\eqa where each of
the $\hat{U}$ matrices corresponds to the rectangle and represents
the following product of its triangular constituents:

\bqa U_{j_1\,j_2\,j_3\,j_4}\left(x, \Delta x^\mu \Delta
x^\nu\right) = \left(E_{I,\kappa}^{ \hat{B}_{\mu\nu}[x] \, \Delta
x^\mu \, \Delta x^\nu}\right)_{j_1\,j_2 \,n} {\left(E_{I,\kappa}^{
\hat{B}_{\mu\nu}\left[x\right] \, \Delta x^\mu \, \Delta
x^\nu}\right)^n}_{j_3\,j_4}  \eqa Thus, we have the following
curvature for the two--tensor field:

\bqa F^{ijk}_{\mu\nu\alpha} = \pr_{[\mu}B^{ijk}_{\nu\alpha]}.\eqa
As the result, the corresponding local field theory is free. Again
this is not a surprise for us. Somehow once we consider the local
fields as non--linear variables ($\hat{B}: V^3 \to \C$), the
theory for them itself becomes linear! As we have already
mentioned, however, the 2D holonomy matrices

$$\left(E_{I,\kappa}^{{\int\int}_{\Sigma} \hat{B}_{\mu\nu} \, dx_\mu \,
dx_\nu}\right)_{\dots}^{\dots}$$ include interactions due to the
non--linear nature of the $B$--field  and especially of the
background $\hat{I}$. It is worth mentioning at this point that in
\cite{Dolotin:1999hk} the two--tensor field was considered to
carry two color indices, i.e. to be a linear field
($B_{\mu\nu}^{ij}\,\Delta x^\mu \, \Delta x^\nu: V\to V$). Then to
organize the triangulation--independent area--ordering with the
use of the standard exponent one had to include the one--form
gauge field ($A$) as well. As the result the theory for such a
couple of fields ($B,A$) becomes interacting due to the presence
of commutators.

Let us clarify these observations. All this idea of the surface
{\it exponent} and the triangulation--independent area--ordering
was invented just to make a local description of the theory of
non--Abelian two--tensor fields\footnote{By ``non--Abelian'' we
mean just those fields which carry color indices.}. But for the
{\bf local} fields everything becomes commutative if we make them
(and especially the background $\hat{I}$) to carry three color
indices. As the result there are no commutators in this case as we
pointed out at the end of the section 2. And once there are no
commutators, then there are no interactions!

It seems that to obtain a non--trivial ordering one has to throw
away a line out of the surface. The corresponding line observable
can be sensitive to the non--trivial ordering. But such a
non--local observable should not be constructed only from the
local $B$--field. As the result, it is non--local theory which can
be non--commutative, while that for local fields is commutative
\cite{Akhmedov:2005zi}. Thus, at this stage the problem is not
that we do not know just how to write the interacting {\bf local}
theory for non--Abelian two--tensor fields, rather we even do not
know the nature of their possible interactions. What can we say?..
Probably this is just the way the nature is and we should be happy
that it is so, because non--linear fields are free and easy to
deal with! The question is whether one can construct anything
non--trivial with the use of such free fields. We are going to
discuss this point right now.

\section{Future directions}

\subsection{Relation to the Yang--Baxter equation}

In this subsection we would like to point out a curious relation
of the above considerations to the Yang--Baxter equations in the
theory of integrable lattice models. The relation is as follows
(see e.g. \cite{Larsson:2002pk}). Consider the surface holonomy
matrix for the cube (see fig. \ref{cube}). The condition of the
zero curvature establishes that the cube holonomy matrix is
trivial:

\bqa \hat{U}\left(x, \Delta x^\mu \Delta x^\nu\right)
\hat{U}\left(x + \Delta x^\nu, \Delta x^\mu \Delta x^\alpha\right)
\, \hat{U}\left(x, \Delta x^\nu \Delta x^\alpha\right) \times
\nonumber \\ \times \hat{U}^{-1}\left(x, \Delta x^\nu \Delta
x^\mu\right) \hat{U}^{-1}\left(x + \Delta x^\mu, \Delta x^\alpha
\Delta x^\nu\right) \, \hat{U}^{-1}\left(x + \Delta x^\alpha,
\Delta x^\nu \Delta x^\alpha\right)=1.\eqa
 This equation is just the Yang--Baxter
equation where the matrices $U_{j_1\,j_2\,j_3\,j_4}\left(x, \Delta
x^\mu \Delta x^\nu\right)$ in \eq{YB} play the role of the
R--matrix.

In this context it is interesting to see whether one can clasify
all possible R--matrices according to all possible choices of
$\left(E_{I,\kappa}^{\hat{B}}\right)_{j_1\,j_2 \,j_3\,j_4}$. We
would like to remind at this point about the star--triangle
equality (which is well in the spirit of the considerations of our
paper) and its relation to the Yang--Baxter equation (see e.g.
\cite{Isaev:2003tk}).

 Apart from other things, these considerations are
important to establish Hamiltonian formalism for
multi--directional evolution. This is the subject of the next
subsection.

\subsection{Relation to the String Field Theory}

Consider for the time being the case when the target space $X$ is
just $R^D$. As well let us take $V = \mathcal{H}$, where
$\mathcal{H}$ is the Hilbert space. Then the indices $i,j$ and $k$
take continuous values and we take them to be $x,y$ and $z$ ---
the coordinates in $X$. Let us take as the base of our fiber
bundle the world--sheet (space of $\sigma$ and $\tau$) rather than
the target space $X$. I.e. in this case the $B_{\mu\nu}$--field is
just the  density $\hat{B}_{\sigma\tau}(\sigma,\tau) =
\hat{B}(\sigma,\tau)$ with three color indices: $\hat{B}: {\cal
H}^3 \to \C$.

  As the result the observables of the theory are as follows:

\bqa U\left[\Sigma^{\phantom{\frac12}}(\gamma_1, \dots,
\gamma_L)\right]^{y^{(1)}(t_1)\left| y^{(2)}(t_2)\right| \dots
\left| y^{(m)}(t_m)\right.}_{x^{(1)}(s_1)\left|
x^{(2)}(s_2)\right| \dots \left| x^{(n)}(s_n)\right.} = \nonumber
\\ =  \left\langle y^{(1)}(t_1)\right| \left\langle
y^{(2)}(t_2)\right| \dots \left\langle y^{(m)}(t_m)\right|
E_{g,I,\kappa}^{-\int\int_{\Sigma} d\sigma d\tau
\hat{B}(\sigma,\tau)} \left| x^{(1)}(s_1) \right\rangle \left|
x^{(2)}(s_2) \right\rangle \dots \left| x^{(n)}(s_n)
\right\rangle.\label{obser}\eqa Here $|x(s)\rangle = \prod_s |
x_s\rangle$ and $|x_s\rangle$ is the standard coherent state in
quantum mechanics. The observables (\ref{obser}) represent {\bf
interacting} string theory amplitudes for the proper choice of
$\hat{I}$ and $\hat{\kappa}$ and $\hat{B}(\sigma,\tau)$. This is a
much better situation with respect to the standard String Field
Theory Hamiltonian quantization \cite{Witten:1985cc}, where one
starts with the free string theory amplitude:

\bqa U({\rm cylinder}) = \left\langle x(\sigma)\right| e^{- H \,
T} \left| y(\sigma)\right\rangle, \quad {\rm where} \quad H =
\int_0^{2\pi} d\sigma \left( - \frac{\delta^2}{\delta x^2(\sigma)}
+ \frac{\left[\pr_\sigma x(\sigma)\right]^2}{2\pi}\right).\eqa
Only after that one includes interactions as perturbative
expansion in the cubic interaction over the quadratic
background\footnote{We do not pay attention to the constraints and
ghosts at this stage.}. The latter formalism suffers from the well
known background dependence, apart from many other things. We see,
however, that this problem does not appear in our formalism which
is intrinsically cubic, i.e. does not have to be expanded over a
quadratic part! To make this statement completely rigorous one has
to establish the following relation explicitly:

\bqa\left\langle x(\sigma)\right| E_{I,\kappa}^{-\int\int_{\Sigma}
d\sigma d\tau \hat{B}(\sigma,\tau)} \left| y(\sigma) \right\rangle
= \left\langle x(\sigma)\right| e^{- H \, T} \left|
y(\sigma)\right\rangle.\eqa I.e. for given $H$ one has to find
explicit values for $\hat{I}$, $\hat{\kappa}$ and
$\hat{B}(\sigma,\tau)$ in this equation for the arbitrary choice
of $x(\sigma)$ and $y(\sigma)$. Which is going to be done in a
separate publication.

It is worth mentioning at this point that the considerations of
this section along with those of the subsection 3.3. bring us very
close to the subject of simplicial strings and open/closed string
duality in the spirit of
\cite{Akhmedov:2004yb},\cite{Akhmedov:2005mr},
\cite{Gopakumar:2003ns} and \cite{Gaiotto:2003yb}. As well we
think that these considerations will elaborate on the nature of
the integrability in the large $N_c$ Yang--Mills theory
\cite{Minahan:2002ve}.

\section{Conclusions and Acknowledgments}

Thus, we have obtained local field theory description of the
non--Abelian two--tensor fields. Due to the triviality of the
corresponding homotopy groups in more than one--dimension, there
are no commutator terms for the local two--tensor fields. Once
there are no commutators, then there are no interactions, i.e. we
obtain a free field theory for the local non--Abelian two--tensor
fields. (We call them non--Abelian because they carry color
indices.) However, this does not mean that theory for them does
not have any non--trivial content.  In fact, because the
background $I_{ijk}$ and the field $B_{\mu\nu}^{ijk}$ carry three
color indices the 2D holonomy matrices in this theory describe
non--trivial maps between fibers of the bundles over loop spaces.

Due to the latter fact we find a promising link of our
constructions to the background independent formulation of String
Field Theory. At the same time our theory is linked with the
R--matrix and the integrability in the lattice models. These
observations all together form a knot which remains to be
disentangled for the better understanding of all mentioned
subjects.

Apart from the future directions listed in the main body of the
text we would like to bring attention to another important
problem. We would like to find a reduction of the non--Abelian
two--tensor fields to the non--Abelian one--form gauge fields. If
the surface, over which the area--ordering is done, is of the
cylindrical topology, it can be degenerated into a curve in the
target space. In this case we can expect to reproduce somehow the
path ordering of a one--form gauge field over the curve. However,
this problem is much more complicated in comparison with the one
presented in the subsection 5.2. In fact, if we deal with a
one--form gauge field $A_\mu$, $\mu=1,\dots,D$ for $D \geq 2$ the
ordering over the curve is non--trivial. And the question is how
to reproduce such an ordering from the trivial one for the
two--tensor fields. This is a problem for a separate study.

I would like to acknowledge valuable discussions with U.Schreiber,
T.Tada, T.Pilling, N.Amburg, D.Vasiliev, A.Losev, A.Rosly and
especially with A.Gerasimov, G.Sharigin, V.Dolotin, S.Loktev and
A.Morozov. I would like to thank Tada san and Ishimoto san for
very worm hospitality at RIKEN, Tokyo, where this work was
finished. This work was done under the partial support of grants
RFBR 04-01-00646, INTAS 03--51--5460 and the Grant from the
President of Russian Federation MK--2097.2004.2.


\begin{thebibliography}{99}

\bibitem{Baez:2004in}
  J.~Baez and U.~Schreiber,
  arXiv:hep-th/0412325.

\bibitem{Baez:2005sn}
  J.~C.~Baez, D.~Stevenson, A.~S.~Crans and U.~Schreiber,
  arXiv:math.qa/0504123.

\bibitem{Akhmedov:2005tn}
  E.~T.~Akhmedov,
  ``Towards the theory of non-Abelian tensor fields. I,''
  arXiv:hep-th/0503234, to appear in Teor. Mat. Fiz..

\bibitem{Verlinde:1988sn}
  E.~Verlinde,
  Nucl.\ Phys.\ B {\bf 300}, 360 (1988).

\bibitem{Fukuma:1992hy}
  M.~Fukuma, S.~Hosono and H.~Kawai,
  Commun.\ Math.\ Phys.\  {\bf 161}, 157 (1994)
  [arXiv:hep-th/9212154].


\bibitem{Akhmedov:2005zi}
  E.~T.~Akhmedov, V.~Dolotin and A.~Morozov,
  ``Comment on the surface exponential for tensor fields,''
  arXiv:hep-th/0504160; to appear in JETF Lett..

\bibitem{Kontsevich:1992ti}
  M.~Kontsevich,
  Commun.\ Math.\ Phys.\  {\bf 147}, 1 (1992).

\bibitem{Dolotin:1999hk}
  V.~Dolotin,
  arXiv:math.gt/9904026.

\bibitem{Larsson:2002pk}
  T.~A.~Larsson,
  arXiv:math-ph/0205017.

\bibitem{Isaev:2003tk}
  A.~P.~Isaev,
  Nucl.\ Phys.\ B {\bf 662}, 461 (2003)
  [arXiv:hep-th/0303056].

\bibitem{Witten:1985cc}
  E.~Witten,
  Nucl.\ Phys.\ B {\bf 268}, 253 (1986).

\bibitem{Akhmedov:2004yb}
E.~T.~Akhmedov,
JETP Lett.\  {\bf 80}, 218 (2004) [Pisma Zh.\ Eksp.\ Teor.\ Fiz.\
{\bf 80}, 247 (2004)] [arXiv:hep-th/0407018].

\bibitem{Akhmedov:2005mr}
  E.~T.~Akhmedov,
  arXiv:hep-th/0502174.   [Pisma Zh.\ Eksp.\ Teor.\ Fiz.\
{\bf 81}, 445 (2005)]

\bibitem{Gopakumar:2003ns}
  R.~Gopakumar,
  Phys.\ Rev.\ D {\bf 70}, 025009 (2004)
  [arXiv:hep-th/0308184];
  R.~Gopakumar,
  Phys.\ Rev.\ D {\bf 70}, 025010 (2004)
  [arXiv:hep-th/0402063];
  R.~Gopakumar,
  arXiv:hep-th/0504229.

\bibitem{Gaiotto:2003yb}
  D.~Gaiotto and L.~Rastelli,
  arXiv:hep-th/0312196.

\bibitem{Minahan:2002ve}
  J.~A.~Minahan and K.~Zarembo,
  JHEP {\bf 0303}, 013 (2003)
  [arXiv:hep-th/0212208].

\end{thebibliography}
\end{document}